\documentclass[a4paper]{jpconf}

\usepackage{citesort}

\usepackage{multirow}
\usepackage{graphicx}
\usepackage{dcolumn}
\usepackage{bm}
\usepackage{soul}
\usepackage{color}

\begin{document}
\title{Exploring the neutron dripline two neutrons at a time: The first observations of the $^{26}$O and $^{16}$Be ground state resonances}

\author{Z. Kohley$^{1,*}$, A.~Spyrou$^{1,2}$, E.~Lunderberg$^3$, P.~A.~DeYoung$^3$, H.~Attanayake$^4$, T.~Baumann$^1$, D.~Bazin$^1$, B.~A.~Brown$^{1,2}$ G.~Christian$^{1,2}$, D.~Divaratne$^4$, S.~M.~Grimes$^4$, A.~Haagsma$^4$, J.~E.~Finck$^5$, N.~Frank$^6$, B.~Luther$^7$, S.~Mosby$^{1,2}$, T.~Nagi$^3$, G.~F.~Peaslee$^3$, W.~A.~Peters$^8$, A.~Schiller$^4$, J.~K.~Smith$^{1,2}$, J.~Snyder$^{1,2}$, M.~J.~Strongman$^{1,2}$, M.~Thoennessen$^{1,2}$, and A.~Volya$^9$}

\address{$^1$National Superconducting Cyclotron Laboratory, Michigan State University, East Lansing, Michigan 48824, USA}
\address{$^2$Department of Physics $\&$ Astronomy, Michigan State University, East Lansing, Michigan 48824, USA}
\address{$^3$Department of Physics, Hope College, Holland, Michigan 49423, USA}
\address{$^4$Department of Physics $\&$ Astronomy, Ohio University, Athens, Ohio 45701, USA}
\address{$^5$Department of Physics, Central Michigan University, Mt. Pleasant, Michigan 48859, USA}
\address{$^6$Department of Physics $\&$ Astronomy, Augustana College, Rock Island, Illinois 61201, USA}
\address{$^7$Department of Physics, Concordia College, Moorhead, Minnesota 56562, USA}
\address{$^8$Department of Physics $\&$ Astronomy, Rutgers University, Piscataway, New Jersey 08854, USA}
\address{$^9$Department of Physics, Florida State University, Tallahasee, Florida 32306, USA}

\ead{$^{*}$zkohley@gmail.com}

\begin{abstract}
The two-neutron unbound ground state resonances of $^{26}$O and $^{16}$Be were populated using one-proton knockout reactions from $^{27}$F and $^{17}$B beams.  A coincidence measurement of 3-body system (fragment + n + n) allowed for the decay energy of the unbound nuclei to be reconstructed.  A low energy resonance, $<$~200~keV, was observed for the first time in the $^{24}$O~+~n~+~n system and assigned to the ground state of $^{26}$O.  The $^{16}$Be ground state resonance was observed at 1.35~MeV.  The 3-body correlations of the $^{14}$Be + n +~n system were compared to simulations of a phase-space, sequential, and dineutron decay.  The strong correlations in the n-n system from the experimental data could only be reproduced by the dineutron decay simulation providing the first evidence for a dineutron-like decay.
\end{abstract}

\section{Introduction}
The addition or removal of neutrons from stable isotopes has been shown in many cases to drastically alter the structure of the nucleus~\cite{BAUMANN12,BROWN01,Tho04}.  In moving towards the neutron dripline the typical shell gaps observed in stable nuclei have been shown to disappear and new magic numbers are observed.  For example, in the dripline nucleus $^{24}$O a large energy difference was measured between the $\nu 1s_{1/2}$ and $\nu 0d_{3/2}$ orbitals indicating a new magic number at $N$ = 16~\cite{Hof08,Hof09}.  Determining how the shell structure changes for nuclei with extreme neutron-to-proton ratios is critical for providing stringent constraints on theoretical calculations.  Such measurements have already been used to demonstrate the importance of three-nucleon forces in describing the properties of very neutron-rich nuclei~\cite{Ots10}.

\par
Radioactive-ion beam facilities have made it possible to produce and study nuclei at and beyond the neutron dripline.  The neutron dripline is defined experimentally only up to the oxygen isotopes, where $^{24}$O is known to be the last bound isotope~\cite{Gui90,Tar97,Sch05}.  The fluorine dripline is known to extend at least to $^{31}$F but heavier isotopes have not been confirmed bound or unbound.  A striking difference is observed in comparing the location of the neutron dripline for the oxygen and fluorine isotopes.  While the 8 protons in oxygen are able to bind up to 16 neutrons, the addition of a single proton enables at least 22 neutrons to be bound.  No published theoretical calculation has been able to simultaneously reproduce both the oxygen and fluorine driplines.  Thus, measuring the properties of the unbound oxygen isotopes can constrain the theoretical calculations and help offer insight into the structure of the neutron dripline.  Predictions of the two-neutron separation energy for $^{26}$O have shown large variations~\cite{Ots10,Shu11,Brown06,Nak08,Volya06}.  An experimental measurement of the two-neutron separation energy would constrain the theoretical calculations and, thus, provide new insights into the changing structure of the neutron dripline.

\par
These exotic dripline nuclei can also exhibit unique types of decay which can provide additional insight into the structure of the nucleus.  Goldanksy was the first to propose that the simultaneous emission of two protons could be observed given a scenario in which the intermediate state was positioned above the initial state~\cite{Gold60}.  Similarly, if the same conditions are found at the neutron dripline the direct emission of two neutrons would be expected.  This scenario exists for the two-neutron unbound $^{16}$Be, which should decay through the direct emission of two neutrons to $^{14}$Be. The limiting cases for the simultaneous two-neutron emission would be a phase-space decay where the neutrons exhibit no correlation or the emission of a dineutron cluster.  The more exotic dineutron decay is of great interest as multiple studies have shown indirect evidence for the presence of a dineutron in neutron-rich halo nuclei, such as $^{6}$He and $^{11}$Li~\cite{Seth91,Ogan99,Ter98,Gal02,Khoa04,NAKAMURA06,Oga94}.  However, a direct measurement of a dineutron-like emission has not yet been observed.  Two-neutron unbound nuclei provide an excellent opportunity to observe a dineutron type decay since the emission occurs directly from the ground state in comparison to measurements of the two-neutron decay from bound nuclei where the system must be excited to overcome the two-neutron separation energy and, thus, the correlations may be altered relative to the ground state.

In the following, we report on the first observations of the ground state resonances of $^{26}$O and $^{16}$Be.  These results were first reported in Refs.~\cite{LUN12} and~\cite{SPYROU12} for $^{26}$O and $^{16}$Be, respectively.


\section{Experiment}
The radioactive $^{27}$F and $^{17}$B beams were produced at the Coupled Cyclotron Facility at the National Superconducting Cyclotron Laboratory at Michigan State University.  Two separate experiments were completed to measure the ground state resonances.  In each case, a stable primary beam was fragmented on a thick Be production target producing a wide variety of isotopes.  The A1900 fragment separator was used to select the desired secondary beam through the removal of the other reaction products and primary beam.  The relevant information about each experiment is presented in Table~\ref{t:Exp}.

\begin{table*}
\begin{center}
\caption{Details of the experiments used to produce the unbound $^{26}$O and $^{16}$Be nuclei.}
\setlength{\extrarowheight}{1.5pt}
\setlength{\tabcolsep}{10pt}
\begin{tabular}{c c c c c  c}
\hline
\hline
Unbound   &Primary  &Secondary                 &Reaction      &Secondary    &\multirow{2}{*}{Ref.}\\
nucleus    &beam    &beam                           &target     &beam rate &\\
\hline
\multirow{2}{*}{$^{26}$O}   &140 MeV/u       &82 MeV/u                   &705 mg/cm$^{2}$    &\multirow{2}{*}{14 pps} &\multirow{2}{*}{\cite{LUN12}}\\
                             &$^{48}$Ca        &$^{27}$F                        &$^{9}$Be                  & &\\
\hline
\multirow{2}{*}{$^{16}$Be}   &120 MeV/u       &53 MeV/u                   &470 mg/cm$^{2}$    &\multirow{2}{*}{250 pps} &\multirow{2}{*}{\cite{SPYROU12}}\\
                             &$^{22}$Ne        &$^{17}$B                        &$^{9}$Be                  & &\\
\hline
\hline
\end{tabular}
\label{t:Exp}
\end{center}
\end{table*}

\par
Measurement of the three-body decay (frag + n + n) required coincident detection of both neutrons and the residual fragment, which was accomplished using the modular neutron array (MoNA)~\cite{MONA03,MONA05} and the large gap 4~Tm sweeper magnet~\cite{SWEEPER}.  The large area neutron detector, MoNA, is composed of 144 plastic scintillator bars.  Light guides and photomultiplier tubes are attached to the ends of each bar for detection of the light produced from the interaction of the neutron(s).  Each bar is 200 cm $\times$ 10 cm $\times$ 10 cm and are typically configured into 9 walls, each 16 bars high.  The time-of-flight (ToF) of the neutrons to MoNA is measured with respect to a scintillator placed in front of the target.  The angle and energy of a neutron can be determined from the interaction point in the scintillator bar and the ToF, respectively.  The sweeper magnet bends the charged particles $\sim$43$^{\circ}$ into a suite of position sensitive charged-particle detectors.  Both the particle identification and kinematical properties of the charged fragments can be determined from the sweeper magnet detectors (see Ref.~\cite{Chr12} for more details).

\par
From the measured energy and angle of the neutrons and charged particle, the invariant mass of $^{26}$O ($M_{^{26}\mathrm{O}}$) and $^{16}$Be ($M_{^{16}\mathrm{Be}}$) could be calculated.  The 3-body decay energy could then be calculated as $E_{\mathrm{decay}} = M_{^{26}\mathrm{O}} - M_{^{24}\mathrm{O}} - 2M_{\mathrm{n}}$ in the $^{26}$O case, where $M_{^{24}\mathrm{O}}$ is the rest mass of $^{24}$O and $M_{\mathrm{n}}$ is the neutron mass.  In the two-neutron decays, the 2-body decay energy can also be examined and is calculated, for the $^{26}$O example, as $E_{\mathrm{decay}} = M_{^{25}\mathrm{O}} - M_{^{24}\mathrm{O}} - M_{\mathrm{n}}$, where $M_{^{25}\mathrm{O}}$ is the invariant mass calculated from the charged particle and the first detected neutron in MoNA.

\section{Monte Carlo Simulation}
A detailed Monte Carlo simulation has been developed to simulate the production and decay of the unbound nuclei.  While a general discussion of the Monte Carlo simulation is provided in Ref.~\cite{Koh12}, the simulation of the neutron interactions in MoNA is of particular importance.  In measuring the two-neutron decays, it is imperative that the simulation be able to accurately reproduce the multiple scattering of neutrons within the array.  A single neutron can scatter twice producing a false 2n signal in comparison to two neutrons each interacting within the array providing a true 2n signal.  The simulation must be able to correctly reproduce the ratio of false to true 2n events in order to allow for the resonance parameters of the experimental data to be extracted.

\par
The MoNA simulation is built upon the \texttt{\sc Geant4} framework~\cite{GEANT4,GEANT42}, which allows for the tracking of each neutron interaction throughout the array.  In order to validate the accuracy of the Monte Carlo simulation, a ``clean'' experimental data set is required for comparison. Ideally this would be a case where only a single neutron is impinging on MoNA.  The measurement of the ground state of $^{16}$B by Spyrou \textit{et al.}~\cite{SPYROU10} was chosen for this reason.  A one-proton knockout from $^{17}$C produced the unbound $^{16}$B, which immediately decayed into $^{15}$B + n.  For each detected $^{15}$B fragment there should only be a single neutron in coincidence.  Therefore, any multiple interaction events observed in MoNA must correspond to the multiple scattering of the single neutron and should allow for a detailed comparison with the simulation.

\par
\begin{figure}
\begin{center}
\includegraphics[width=0.56\textwidth]{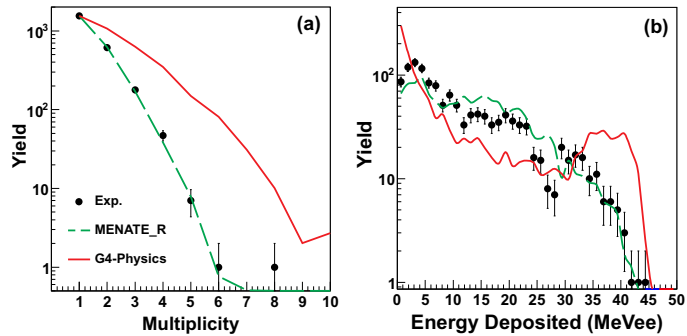}
\caption{\label{f:Menate} (Color online) The experimental (a) multiplicity and (b) deposited energy distributions are compared to the Monte Carlo simulation using the \textsc{menate\_r} and G4-Physics classes.  The simulated multiplicity distributions were scaled to match the number of multiplicity~=~1 experimental events.  The simulated deposited energy distributions were normalized to the total area of the experimental distribution.}
\end{center}
\end{figure}

In Fig.~\ref{f:Menate} the multiplicity and energy deposited distributions from hits in coincidence with $^{15}$B are shown from the experiment in comparison to the \texttt{\sc Geant4} simulation.  The standard physics classes, referred to as G4-Physics, included in the \texttt{\sc Geant4} package are unable to reproduce either of the distributions.  This was found to be related to the treatment of the inelastic neutron-carbon reaction channels~\cite{Koh12}.  In the G4-Physics simulation, the \texttt{\sc Geant4} cascade model is used to determine how the different possible inelastic channels are partitioned~\cite{G4Physics}.  In comparison to the G4-Physics, a custom neutron interaction model, referred to as \textsc{menate\_r}~\cite{menateR,MENATE}, was incorporated into the \texttt{\sc Geant4} framework.  While the neutron(s) was still transported using the typical \texttt{\sc Geant4} methods, the \textsc{menate\_r} class determined how each interaction was treated.   The \textsc{menate\_r} code determines the partitioning of each of the different inelastic neutron-carbon reaction channels based on a compilation of experimental cross sections.  Thus, the relative rates for each reaction channel should be accurately reproduced.  A drastic improvement in the accuracy of the simulation is observed with the inclusion of \textsc{menate\_r} within \texttt{\sc Geant4}~\cite{Koh12}, as shown in Fig.~\ref{f:Menate}.

\par
\begin{figure}
\begin{center}
\includegraphics[width=0.65\textwidth,height=0.27\textheight, trim= 11.0mm 0.0mm 7.0cm 0.0mm]{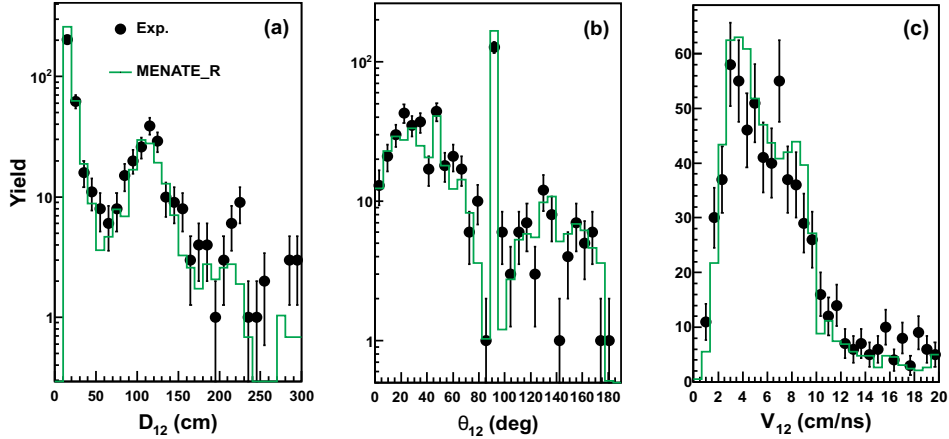}
\caption{\label{f:cuts} (Color online) Relative (a) distance, (b) angle, and (c) velocity between the first and second interaction in MoNA from the experiment is compared with the Monte Carlo simulation with the \textsc{menate\_r} neutron interaction model.  The simulated distributions were normalized to the total area of the experimental distributions.}
\end{center}
\end{figure}

As mentioned above, the two-neutron measurements are dependent on the ability to separate true and false 2n events.  In both the $^{26}$O and $^{16}$Be studies, causality cuts placed on the relative distance and velocity between the first and second hits in MoNA were used to remove false 2n events.  In Fig.~\ref{f:cuts} the relative distance ($D_{12}$), angle ($\theta_{12}$), and velocity ($V_{12}$) are shown from the experimental $^{16}$B decay data~\cite{SPYROU10} and the simulation using \textsc{menate\_r}.  Since each event in the experiment should only have one neutron in coincidence, the distributions, which require two hits in MoNA, are representative of multiple scattering or false 2n events.  It is clear that false 2n signals are largely correlated with a small $D_{12}$, roughly less than 30~cm, and a $V_{12}$ less than about 10~cm/ns (which is equivalent to the beam velocity).  This demonstrates the basis for the applied 2n casuality cuts used in the analysis of the $^{26}$O and $^{16}$Be.  The most important aspect is the excellent agreement shown between the experimental data and the Monte Carlo simulation.  This indicates that the simulation is able to accurately reproduce the multiple scatterings of the neutrons within MoNA and should, therefore, be able to reproduce the effects of applying the 2n causality cuts.

\section{$^{26}$O}

As mentioned above, the unbound ground state of $^{26}$O was searched for using a one-proton knockout from a $^{27}$F beam and measuring the triple coincidence of the charged particle and two neutrons.  While theoretical predictions have varied significantly on the prediction of the ground state position, the continuum shell model (CSM) has shown excellent reproduction of the measured properties of the oxygen isotopes and has predicted the $^{26}$O ground state to be unbound by only 21~keV~\cite{Volya06,Vol05}.  Thus, the presence of low energy neutrons would be an immediate indication for the $^{26}$O ground state resonance.
\par
\begin{figure}
\begin{center}
\includegraphics[width=0.85\textwidth]{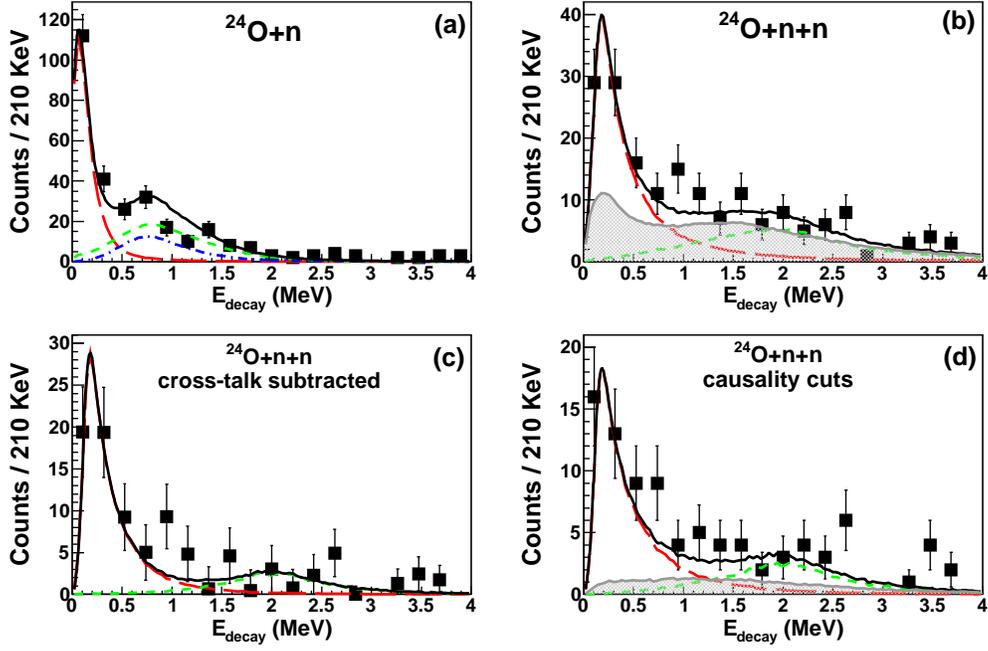}
\caption{\label{f:26O_decay} (a) Decay energy spectrum of the two-body system, $^{24}$O + n.  (b) Three-body decay energy spectrum for the $^{26}$O system.  (c) and (d) show the three-body decay energy spectrum after the subtraction of the simulated false 2n events and after applying the causality cuts, respectively.  The experimental data are shown as solid black circles.  The three decay channels used in the Monte Carlo simulation are described in the text and shown in Fig.~\ref{f:26O_fit}.  The grey-shaded region in panels (b) and (d) represents the simulated false 2n component in the spectra.}
\end{center}
\end{figure}

In Fig.~\ref{f:26O_decay}(a) the two-body, $^{24}$O + n, decay energy is presented.  A previous measurement of the $^{25}$O ground state, populated using a $^{26}$F beam, reported a resonance at 770$^{+20}_{-10}$~keV~\cite{Hof08}. As shown, the experimental $E_{decay}$ spectrum does show an increase around 700~keV, which is likely associated with the population of the $^{25}$O ground state.  However, more interesting is the strong increase in $E_{decay}$ below 400~keV, which is the first indication that the $^{26}$O resonance has been observed.  Fig.~\ref{f:26O_decay}(b) shows the 3-body decay energy, where the $^{26}$O is reconstructed from the charged particle and the first two hits in MoNA without causality cuts applied.  Again, a clear increase in the decay energy is observed below 500~keV indicating the presence of a low-energy resonance in the $^{26}$O system.

\par
\begin{figure}
\begin{center}
\includegraphics[width=0.45\textwidth]{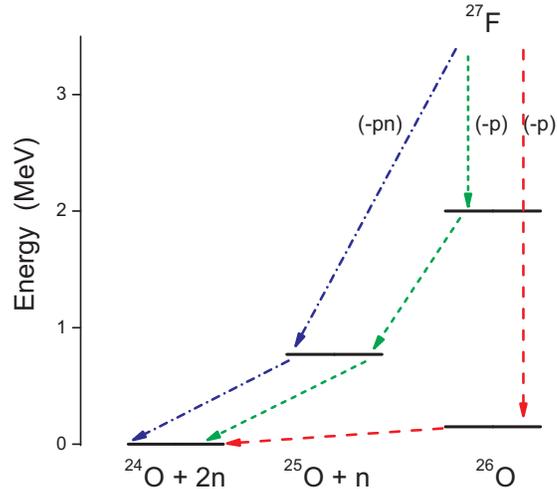}
\caption{\label{f:26O_fit} Energy and decay level diagram for $^{26}$O decaying to $^{24}$O.  The three decay channels used in the Monte Carlo simulation are shown.  The position of the $^{25}$O ground state was taken from the experimental measurement of Hoffman \emph{et al.}~\cite{Hof08} and the $^{26}$O energy levels are from the CSM calculations~\cite{Volya06,Vol05}.}
\end{center}
\end{figure}

The Monte Carlo simulation was used to simultaneously fit Figs.~\ref{f:26O_decay}(a) and \ref{f:26O_decay}(b).  Three decay channels were considered in the Monte Carlo simulation as shown in Fig.~\ref{f:26O_fit}.  The one-proton knockout populating the ground state resonance of $^{26}$O (long-dashed red line) was modeled using a Breit-Wigner lineshape and assuming the two neutrons were emitted simultaneously with a phase space distribution.  The population of the first excited state (short-dashed green line) was also considered and simulated as a sequential decay through the ground state resonance of $^{25}$O.  Lastly, the $^{25}$O ground state (dot-dashed blue line) could also be directly populated through a pn-removal from the $^{27}$F beam.

\par
In the fit of the experimental data, the amplitudes of each decay channel were allowed to vary freely.  However, since the data are not sensitive to the parameters of the possible contributions to the high-energy continuum the resonance parameters of the sequential decay from the first excited state and the direct population of the $^{25}$O ground state were kept constant.  The $^{25}$O ground state resonance was simulated with an energy-dependent Breit Wigner with $E$ = 770~keV, $\Gamma$ = 172~keV, and $L$ = 2 following the results of Ref.~\cite{Hof08}.  The first excited state of $^{26}$O was also represented by a Breit-Wigner with $E$ = 2.0~MeV, taken from the CSM calculations~\cite{Volya06,Vol05}, and $\Gamma$ = 200~keV.  The $^{26}$O ground state resonance was modeled as a $L$ = 2 energy-dependent Breit-Wigner and both the resonance energy and width were free parameters in the fit.

\par
As shown in the top panels of Fig.~\ref{f:26O_decay}, the sum of the three components in the Monte Carlo simulation (solid black line) were able to reproduce the experimental decay energy spectra.  Through minimizing the $\chi^{2}$ of the fit the $^{26}$O ground state resonance energy was determined to be at $E$~=~150$^{+50}_{-150}$~keV and was insensitive to the width of the resonance.   The $^{26}$O ground state resonance (red long-dashed line) is shown in Fig.~\ref{f:26O_decay} with $E$~=~150~keV and $\Gamma$~=~5~keV.

\par
The grey shaded region in the 3-body decay energy (Fig.~\ref{f:26O_decay}(b)) represents the false 2n component estimated by the Monte Carlo simulation.  The false 2n events consist of a significant portion of the spectrum, yet the $^{26}$O ground state resonance appears to consist of mostly true 2n events.  Removal or reduction of the false 2n events can provide additional evidence for the presence of the ground state resonance.  Two methods were used for this: (1) the simulated false 2n events were subtracted from the experimental 3-body spectrum as shown in Fig.~\ref{f:26O_decay}(c) and (2) the causality cuts were applied to the data as shown in Fig.~\ref{f:26O_decay}(d).  After subtracting the simulated false 2n events, the spectrum contains only the ground state resonance and a background level of counts at higher decay energies.  Applying causality cuts of $D_{12} > 25$~cm and $V_{12} > 7$~cm/ns to both the simulated and experimental data shows a reduction in the false 2n events by a factor of 3 with the ground state resonance observed well above the false 2n component.  The persistence of the low decay energy peak after the removal and reduction of the false 2n events demonstrates strong evidence for the observation of the $^{26}$O ground state resonance.

\section{$^{16}$Be}

\begin{figure}
\begin{center}
\includegraphics[width=0.45\textwidth]{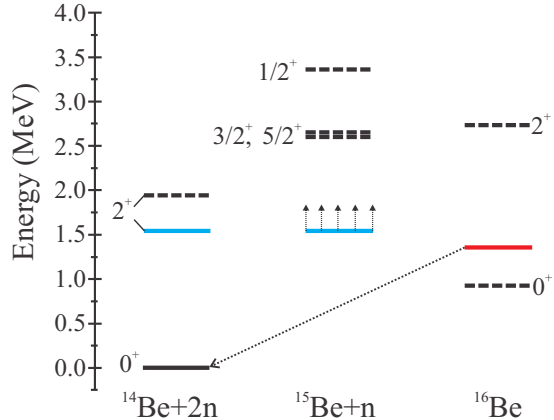}
\caption{\label{f:16Be_level} Energy and decay level diagram for $^{16}$Be.  The solid blue lines represent previous experimental observations of the first excited state in $^{14}$Be~\cite{Sug07} and a lower limit on the position of the $^{15}$Be ground state~\cite{Spy11}.  The $^{16}$Be ground state measurement from the current work is shown as the solid red line.  The dashed black lines represent shell model calculations using the WBP interaction~\cite{War92}. }
\end{center}
\end{figure}

Shell model calculations, as shown in Fig.~\ref{f:16Be_level}, have predicted that $^{16}$Be is bound with respect to one-neutron decay and unbound with respect to two-neutron decay~\cite{War92}.  This provides a unique scenario in which the $^{16}$Be should decay through the simultaneous emission of two neutrons.  Experimental measurements (blue solid line) of the first excited state in $^{14}$Be~\cite{Sug07} and a lower-limit on the $^{15}$Be ground state~\cite{Spy11} both indicate that the shell model calculations appear to be reasonable, which suggests that the $^{16}$Be ground state energy should be around 1.0~MeV.

\par
Based on the scenario presented in Fig.~\ref{f:16Be_level}, three modes of decay for the $^{16}$Be can be envisioned.  The two-neutron emission could proceed via sequential decay through the tail of the broad 1/2$^{+}$ $s$-state around 3.5~MeV predicted by the shell model calculations.  The two-neutrons could also be emitted simultaneously in an uncorrelated phase-space decay or as a correlated pair in a dineutron-like decay.  Examination of the correlations in the 3-body decay can provide insight into the different decay modes.

\par
\begin{figure}
\begin{center}
\includegraphics[width=0.75\textwidth]{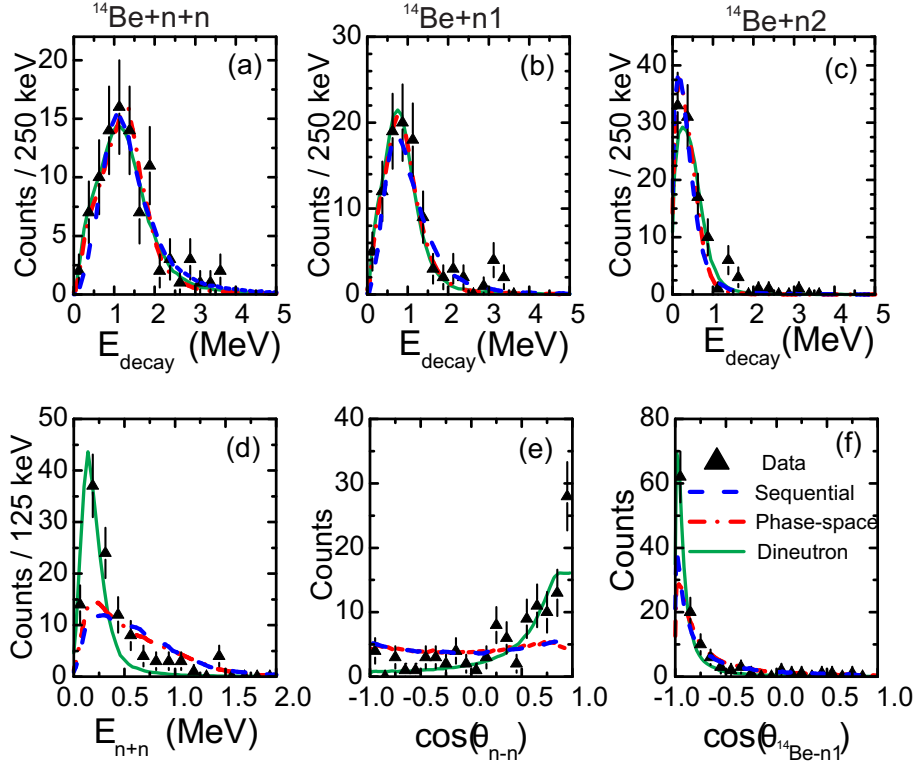}
\caption{\label{f:16Be_exper} The three-body decay energy spectrum for $^{16}$Be is presented in panel (a).  Panels (b) and (c) show the two-body decay energy spectra in which the first and second time-ordered neutron was used, respectively.  Panel (d) shows the decay energy of the n-n system, panel (e) shows the cosine of the opening angle between the two neutrons, and panel (f) shows the angle between the first neutron and the $^{14}$Be fragment.  The causality cuts discussed in the text have been applied to all the spectra, panels (a)-(f). }
\end{center}
\end{figure}

The $^{16}$Be ground state was observed through the coincidence measurement of the $^{14}$Be~+~n~+~n system.  The 3-body decay energy spectrum is shown in Fig.~\ref{f:16Be_exper}(a) with causality cuts of $D_{12} > 50$~cm and $V_{12} > 11$~cm/ns applied.  A clear peak is observed in the 3-body decay energy spectrum around 1.5~MeV indicating the presence of the $^{16}$Be ground state resonance.  Additionally, the 2-body decay energy spectra are shown in panels (b) and (c) using the first (n1) and second (n2) time ordered neutrons measured in MoNA.

\par
The three different decay modes of $^{16}$Be were simulated within the Monte Carlo simulation for comparison with the experimental data.  In each case a Breit Wigner lineshape was used to represent the $^{16}$Be ground state resonance.  The correlations and partitioning of the energy between the emitted neutrons were then determined following the decay calculation.  The sequential decay was calculated following the formulism presented in Ref.~\cite{Volya06} for the emission of two $l = 0$ neutrons with an intermediate $s$-state at about 3.5~MeV.  The simultaneous phase-space emission was simulated using the calculation of Ref.~\cite{PHASESPACE} .  In the dineutron simulation, the two neutrons are emitted together as a cluster and then immediately break-up.  The break-up energy of the n-n system was calculated using a $s$-wave line shape~\cite{Ber98} based on the free space n-n scattering length of -18.7~fm~\cite{Gon06} and a pairing energy of 0.87~MeV~\cite{Ber93}.  The kinetic emission energy of the dineutron cluster is the difference in the total energy of the 3-body system determined from the ground state Breit Wigner and the break-up energy of the 2n cluster.

\par
In panels (a), (b), and (c) of Fig.~\ref{f:16Be_exper} the best fit results of the Monte Carlo simulation for each decay mechanism are shown in comparison to the experimental decay energy spectra.  As expected, the decay mechanism is insensitive to the shape of the decay energy spectra and the experimental data are fit well with each decay type.  A $\chi^{2}$ minimization, varying $E$ and $\Gamma$ of the Breit Wigner, was completed for each decay mechanism.  The resulting best fit corresponds to a $^{16}$Be ground state resonance at $E$~=~1.35(10)~MeV with $\Gamma$~=~0.8$^{+0.1}_{-0.2}$~MeV.  As shown in Fig.~\ref{f:16Be_level} (red line), the measured $^{16}$Be ground state is in agreement with the shell model prediction and presents a scenario for the observation of genuine 3-body process.

\par
The correlations in the $^{14}$Be + n + n system are shown in the bottom panels of Fig.~\ref{f:16Be_exper}.  Panel (d) shows the decay energy of the n-n system.  The opening angle between the two neutrons, in the frame of the 3-body system, is shown in panel (e), while the angle between the first neutron and the $^{14}$Be is shown in panel (f).  Unlike the $E_{decay}$ spectra shown in panels (a)-(c), a clear sensitivity to the decay mechanism is observed in the correlations presented in panels (d)-(f).  The sequential and phase-space decay mechanisms both show a broad $E_{n+n}$ distribution and a relatively flat $cos(\theta_{n-n})$ distribution, which do not reproduce the observed correlations from the experimental data.  In comparison, the dineutron decay simulation is able to reproduce the general features of the experimental data showing a peak in the $E_{n+n}$ distribution and a preference for small neutron-neutron opening angles.  Overall, the agreement between the dineutron decay and experimental data implies that a strong dineutron-like correlation exists between the two neutrons emitted in the decay of $^{16}$Be.  While the decay is most likely not a true two-step process (emission of dineutron followed by the dineutron break-up), the presence of the n-n virtual state ($a_{s} = -18.7$~fm) is clearly observed in this genuine 3-body decay process.

\section{Conclusions}

The ground state resonances of the two-neutron unbound $^{26}$O and $^{16}$Be have been observed for the first time.  The large area neutron detector MoNA and 4~Tm sweeper magnet were used to measure the charged fragment and two neutrons in coincidence allowing for the 3-body decay energy to be reconstructed.  The $^{26}$O ground state resonance was observed at $E$~=~150$^{+50}_{-150}$~keV, in agreement with the sophisticated CSM predictions~~\cite{Volya06,Vol05}.  This extends the experimental measurements of the oxygen dripline and has already proven valuable in constraining theoretical calculations~\cite{Hag12}, which have recently shown the need for three-nucleon forces to reproduce the measured $^{26}$O ground state.

\par
The $^{16}$Be ground state resonance was measured at $E$~=~1.35(10)~MeV with a width of 0.8$^{+0.1}_{-0.2}$~MeV.  This placed the $^{16}$Be ground state below the experimental limit on the $^{15}$Be ground state, which produces a scenario for a genuine three-body decay process.  Three decay processes where simulated: (1) sequential decay through the tail of a broad $s$-state, (2) phase-space decay, and (3) emission of dineutron which proceeds to break-up based on the n-n free space scattering length.  Only the dineutron decay simulation was able to reproduce the experimental 3-body correlation observables suggesting that $^{16}$Be decays through the emission of two strongly correlated neutrons.  Since the simulated decay mechanisms represent extreme limits (dineutron versus phase-space), calculations using a more sophisticated 3-body model will be pursued in the future.  Additionally, these results open the exciting prospect of finding other nuclei that exhibit similar signatures for a dineutron-like decay.

\section*{Acknowledgements}
The authors gratefully acknowledge the support of the NSCL operations staff for providing a high quality beam.  This material is based upon work supported by the Department of Energy National Nuclear Security Administration under Award Number DE-NA0000979 and DOE Award number DE-FG02-92ER40750.
\section*{References}


\begin{thebibliography}{10}
\expandafter\ifx\csname url\endcsname\relax
  \def\url#1{{\tt #1}}\fi
\expandafter\ifx\csname urlprefix\endcsname\relax\def\urlprefix{URL }\fi
\providecommand{\eprint}[2][]{\url{#2}}

\bibitem{BAUMANN12}
Baumann T, Spyrou A and Thoennessen M 2012 {\em Rep. Prog. Phys.\/} {\bf 75}
  036301

\bibitem{BROWN01}
Brown B~A 2001 {\em Prog. Part. Nucl. Phys.\/} {\bf 47} 517

\bibitem{Tho04}
Thoennessen M 2004 {\em Rep. Prog. Phys.\/} {\bf 67} 1187

\bibitem{Hof08}
Hoffman C~R {\em et~al.\/} 2008 {\em Phys. Rev. Lett.\/} {\bf 100} 152502

\bibitem{Hof09}
Hoffman C~R {\em et~al.\/} 2009 {\em Phys. Lett. B\/} {\bf 672} 17

\bibitem{Ots10}
Otsuka T, Suzuki T, Holt J~D, Schwenk A and Akaishi Y 2010 {\em Phys. Rev.
  Lett.\/} {\bf 105} 032501

\bibitem{Gui90}
Guillemaud-Mueller D {\em et~al.\/} 1990 {\em Phys. Rev. C\/} {\bf 41} 937

\bibitem{Tar97}
Tarasov O {\em et~al.\/} 1997 {\em Phys. Lett. B\/} {\bf 409} 64

\bibitem{Sch05}
Schiller A, Baumann T, Dietrich J, Kaiser S, Peters W and Thoennessen M 2005
  {\em Phys. Rev. C\/} {\bf 72} 037601

\bibitem{Shu11}
Shukla A, Aberg S and Patra S~K 2011 {\em J. Phys. G: Nucl. Part. Phys.\/} {\bf
  38} 095103

\bibitem{Brown06}
Brown B~A and Richter W~A 2006 {\em Phys. Rev. C\/} {\bf 74} 034315

\bibitem{Nak08}
Nakada H 2008 {\em Phys. Rev. C\/} {\bf 78}

\bibitem{Volya06}
Volya A and Zelevinsky V 2006 {\em Phys. Rev. C\/} {\bf 74} 064314

\bibitem{Gold60}
Goldansky V~I 1960 {\em Nucl. Phys.\/} {\bf 19} 482

\bibitem{Seth91}
Seth K~K and Parker B 1991 {\em Phys. Rev. Lett.\/} {\bf 66} 2448

\bibitem{Ogan99}
Oganessian Y~T, Zagrebaev V~I and Vaagen J~S 1999 {\em Phys. Rev. Lett.\/} {\bf
  82} 4996

\bibitem{Ter98}
Ter-Akopian G {\em et~al.\/} 1998 {\em Phys. Lett. B.\/} {\bf 426} 251

\bibitem{Gal02}
Galanina L~I and Zelenskaya N~S 2002 {\em Phys. Atom. Nucl.\/} {\bf 65} 1282

\bibitem{Khoa04}
Khoa D~T and von Oertzen W 2004 {\em Phys. Lett. B\/} {\bf 595} 193

\bibitem{NAKAMURA06}
Nakamura T {\em et~al.\/} 2006 {\em Phys. Rev. Lett.\/} {\bf 96} 252502

\bibitem{Oga94}
Ogawa Y, Suzuki Y and Yabana K 1994 {\em Nucl. Phys. A\/} {\bf 571} 784

\bibitem{LUN12}
Lunderberg E, DeYoung P~A, Kohley Z, Attanayake H, Baumann T, Bazin D,
  Christian G, Divaratne D, Grimes S~M, Haagsma A, Finck J~E, Frank N {\em
  et~al.\/} 2012 {\em Phys. Rev. Lett.\/} {\bf 108} 142503

\bibitem{SPYROU12}
Spyrou A, Kohley Z, Baumann T, Bazin D, Brown B~A, Chrstian G, DeYoung P~A,
  Finck J~E, Frank N, Lunderberg E, Mosby S, Peters W~A, Schiller A, Smith J~K,
  Synder J, Strongman M~J, Thoennessen M and Volya A 2012 {\em Phys. Rev.
  Lett.\/} {\bf 108} 102501

\bibitem{MONA03}
Luther B, Baumann T, Thoennessen M, Brown J, DeYoung P, Finck J, Hinnefeld J,
  Howes R, Kemper K, Pancella P, Peaslee G, Rogers W and Tabor S 2003 {\em
  Nucl. Instrum. Meth. A\/} {\bf 505} 33

\bibitem{MONA05}
Baumann T {\em et~al.\/} 2005 {\em Nucl. Instrum. Meth. A\/} {\bf 543} 517

\bibitem{SWEEPER}
Bird M~D, Kenney S~J, Toth J, Weijers H~W, DeKamp J~C, Thoennessen M and Zeller
  A~F 2005 {\em IEEE Trans. Appl. Supercond.\/} {\bf 15} 1252

\bibitem{Chr12}
Christian G {\em et~al.\/} 2012 {\em Phys. Rev. C\/} {\bf 85} 034327

\bibitem{Koh12}
Kohley Z, Lunderberg E, DeYoung P~A, Roeder B~T, Baumann T, Christian G, Mosby
  S, Smith J~K, Snyder J, Spyrou A and Thoennessen M 2012 {\em Nucl. Instrum.
  Meth. Phys. Res. A\/} {\bf 682} 59

\bibitem{GEANT4}
Agostinelli S, Allision J, Amako K, Apostolakis J, Araujo H, Arce P, Asai M,
  Axen D, Banerjee S, Barrand G, Behner F, Bellagamba L, Boudreau J {\em
  et~al.\/} 2003 {\em Nucl. Instrum. Meth. A\/} {\bf 506} 250

\bibitem{GEANT42}
Allision J, Amako K, Apostolakis J, Araujo H, Dubios P~A, Asai M, Barrand G,
  Capra R, Chauvie S, Chytracek R, Cirrone G~A~P, Cooperman G {\em et~al.\/}
  2006 {\em IEEE T. Nucl. Sci.\/} {\bf 53} 270

\bibitem{SPYROU10}
Spyrou A, Baumann T, Bazin D, Blanchon G, Bonaccorso A, Breitbach E, Brown J,
  Christian G, DeLine A, DeYoung P~A, Finck J~E, Frank N, Mosby S, Peters W~A,
  Russel A, Schiller A, Strongman M~J and Thoennessen M 2010 {\em Phys. Lett.
  B\/} {\bf 683} 129

\bibitem{G4Physics}
Geant4 {P}hysics {R}eference {M}anual. http://geant4.web.cern.ch. (2011).

\bibitem{menateR}
Roeder B {D}evelopment and validation of neutron detection simulations for
  {EURISOL} {EURISOL} {D}esign {S}tudy, {R}eport:
  [10-25-2008-006-In-beamvalidations.pdf, pp 31-44] (2008),
  www.eurisol.org/site02/physics and instrumentation/

\bibitem{MENATE}
Desesquelles P, Cole A~J, Dauchy A, Giorni A, Heuer D, Lleres A, Morand C,
  Sain-Martin J, Stassi P, Viano J~B, Chambon B, Cheynis B, Drain D and Pastor
  C 1991 {\em Nucl. Instrum. Meth. A\/} {\bf 307} 366

\bibitem{Vol05}
Volya A and Zelevinksy V 2005 {\em Phys. Rev. Lett.\/} {\bf 94} 052501

\bibitem{War92}
Warburton E~K and Brown B~A 1992 {\em Phys. Rev. C\/} {\bf 46} 923

\bibitem{Sug07}
Sugimoto T {\em et~al.\/} 2007 {\em Phys. Lett. B\/} {\bf 654} 160

\bibitem{Spy11}
Spyrou A {\em et~al.\/} 2011 {\em Phys. Rev. C\/} {\bf 84} 044309

\bibitem{PHASESPACE}
James F {CERN,} {Y}ellow {R}eport {N}o. 68-15 (1968).

\bibitem{Ber98}
Bertsch G~F, Hencken K and Esbensen H 1998 {\em Phys. Rev. C\/} {\bf 57} 1366

\bibitem{Gon06}
Trotter D~E~G {\em et~al.\/} 2006 {\em Phys. Rev. C\/} {\bf 72} 034001

\bibitem{Ber93}
Bertulani C~A, Canto L~F and Hussein M~S 1993 {\em Phys. Rep.\/} {\bf 226} 281

\bibitem{Hag12}
Hagen G, Hjorth-Jensen M, Jansen G~R, Machledit R and Papenbrock T 2012 {\em
  Phys. Rev. Lett.\/} {\bf 108} 242501

\end{thebibliography}
\providecommand{\newblock}{}

\smallskip

\end{document}